\documentclass[twocolumn,showpacs,superscriptaddress,preprintnumbers,amsmath,amssymb,pra]{revtex4}
\usepackage{}
\usepackage{mathrsfs}
\usepackage{amsmath}
\usepackage{amssymb}
\usepackage{graphicx}
\usepackage{dcolumn}
\usepackage{bm}
\usepackage{subfigure}
\usepackage{color}
\usepackage{ifpdf}
\usepackage{epstopdf}
\usepackage{CJK}
\usepackage[
  pdfstartview=FitH,%
  bookmarks=true,%
  bookmarksopen=true,%
  breaklinks=true,%
  colorlinks=true,%
  linkcolor=blue,anchorcolor=blue,%
  citecolor=blue,filecolor=blue,%
  menucolor=blue,pagecolor=blue,%
  urlcolor=blue]{hyperref}
\usepackage{float}

\begin{document}
%\begin{CJK*}{GBK}{song}
%\begin{CJK*}{UTF8}{gbsn}

\title{Observation of quantum dynamical oscillations of ultracold atoms in the F and D bands  of an optical lattice}

\author{Zhongkai Wang}
\affiliation{School of Electronics Engineering and Computer Science, Peking University, Beijing 100871, China}
\author{Baoguo Yang}
\affiliation{School of Electronics Engineering and Computer Science, Peking University, Beijing 100871, China}
\author{Dong Hu}
\affiliation{School of Electronics Engineering and Computer Science, Peking University, Beijing 100871, China}
\author{Xuzong Chen}
\affiliation{School of Electronics Engineering and Computer Science, Peking University, Beijing 100871, China}
\author{Hongwei Xiong}
\affiliation{Wilczek Quantum Center, College of Science, Zhejiang University of Technology, Hangzhou 310014, China}
\author{Biao Wu}
\affiliation{Wilczek Quantum Center, College of Science, Zhejiang University of Technology, Hangzhou 310014, China}
\affiliation{International Center for Quantum Materials, School of Physics, Peking University, Beijing 100871, China}
\affiliation{Collaborative Innovation Center of Quantum Matter, Beijing 100871, China}
\author{Xiaoji Zhou}\email{xjzhou@pku.edu.cn}
\affiliation{School of Electronics Engineering and Computer Science, Peking University, Beijing 100871, China}
\date{\today}

\begin{abstract}
  We report the observation of quantum dynamical oscillations of ultracold atomic gases in the F and D bands of a single-well optical lattice. We are able to control  the Bragg reflections at the Brillouin zone edge up to the third order. As a result, we can switch the quantum dynamics from oscillations across both the F and D bands to oscillations only within the F-band. Our capability to observe these remarkable oscillations comes from the innovative non-adiabatic technique which allows us to load ultracold atoms efficiently to the G-band of an optical lattice.
\end{abstract}


\maketitle
%\end{CJK*}

\section{introduction}
There has been a lot of effort both experimentally and theoretically to study the quantum dynamics of ultracold atoms in optical lattices, such as the celebrated Bloch oscillations (BOs)~\cite{Salomon,Morsch,Arimondo,Gustavsson,Fattori,Haller} and the Landau-Zener (LZ) tunneling~\cite{Qian,Biao1,Biao2,Garraway,Arimondo}. These studies have focused on the lowest band as it is hard to load atoms to high bands and then control their quantum dynamics experimentally. Recently, people are pushing the boundary and studying quantum dynamics involving more than one bands. The effort has resulted in the observation of the Bloch-Zener oscillations (BZOs)~\cite{Breid,Mizumoto,Lim,Dreisow,Uehlinger,Ritt,Kling}, where the quantum oscillations are between two Bloch bands and the crossing between these two bands is facilitated by the LZ tunneling.

However, this kind of oscillations within two Bloch bands are very difficult to observe in a simple single-well optical lattice. In this kind of simple lattices, the band gaps are always smaller for higher bands. If atoms can tunnel from the S-band to the P-band, they should also be able to tunnel from the P-band to the D band. As a result, the oscillations would involve uncontrollably many bands~\cite{Slonghi}. To control oscillations within two bands, one has to design lattices with more complex constructions, which include binary superlattice of optical waveguide arrays~\cite{Dreisow}, honeycomb lattice~\cite{Uehlinger}, and mini-band structure~\cite{Ritt,Kling}. Here in this work we demonstrate experimentally that we can initiate and control quantum oscillations with two high excited bands in a simple single-well optical lattice. Our trick is to use the variable external force from the harmonic trap instead of the usual linear external potential.

In this work the Bose-Einstein condensate (BEC) is initially loaded non-adiabatically into the G-band of a one-dimensional single-well optical lattice. It subsequently tunnels to the F-band and begins oscillations within the F and D bands, which are clearly observed in the momentum space. We can control the Bragg reflection between momenta $3\hbar k_L$ and $-3\hbar k_L$, which is at the FBZ edge between the F and D bands, by tuning the optical lattice strength. $k_L$ is the wave vector of the laser forming the lattice. When the Bragg reflection is weak, quantum oscillations crossing the F and D bands are observed. When the reflection is strong, quantum oscillations are observed in only the F-band. When the Bragg reflection is at intermediate strength, we observe the superposition of these two types of quantum oscillations. In our results, the oscillations crossing the two excited bands can last up to 58ms, which is much longer than the BZOs observed in Ref.~\cite{Kling}. During the oscillations, the BEC can be displaced up to $\pm 100 \mu$m (470 lattice sites) in space, which is bigger than the spatial displacement observed in super-Bloch oscillations~\cite{Haller}.

\section{Experiment method to prepare the atoms in the high bands}
In our experiment, a pure BEC of about $1.5\times10^5$ $^{87}$Rb atoms is prepared in a hybrid trap which is formed by overlapping a single-beam optical dipole trap with wave length $1064$nm and a quadrapole magnetic trap. The resulting  harmonic trapping frequencies are $(\omega_x,\omega_y,\omega_z)=2\pi\times(28,55,65)$Hz. After preparing a BEC in the harmonic trap, we use the non-adiabatic shortcut method~\cite{Zhai,Liu,Chen,Campo} to load the BEC in a one-dimensional optical lattice (along the $x$ direction) into the G-band at the quasi-momentum $\hbar q=0$. The optical lattice $V_{0}\cos ^{2}\left(k_L x\right)$ is produced by a standing wave created by two counter-propagating laser beams with the lattice constant $a=\lambda/2=426$nm and $V_{0}$ being the lattice depth.

\begin{figure}
\begin{center}
  \includegraphics[width=0.4\textwidth]{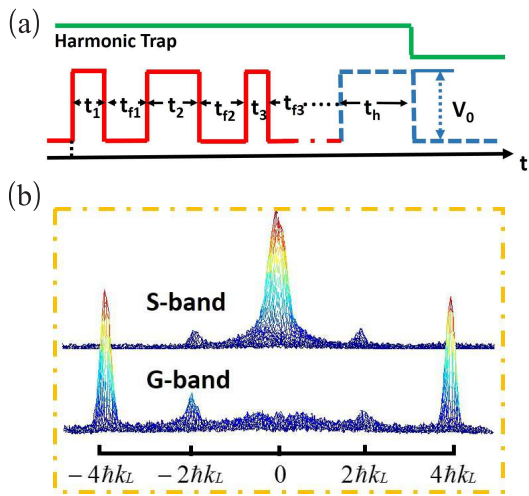}
\end{center}
  \caption{(Color online) (a) The pulse sequence for the loading to high bands. (b) TOF pictures taken after loading atoms into the S and G bands, respectively.}\label{shortcut_method}
\end{figure}

This coherent loading method~\cite{Zhai,Liu} includes a series of designed standing wave pulses shown in Fig.~\ref{shortcut_method}(a). It allows us to non-adiabatically load a BEC from the ground state of the harmonic trap $|\psi_{0}\rangle$ directly into a target state $|\psi_{a}\rangle=|n,q\rangle$. Here $|n,q\rangle$ is the eigenstate of Bloch bands, $n={\rm S},{\rm P},{\rm D}...$ is the band index. The  pulse sequence can be optimized so that the final state is nearly a Bloch state (the fidelity can be over 98\%). This can be verified with the proportion of different momentum components or the fast oscillations of different momentum states. For the F and G bands, $\pm 4\hbar k_L$ momenta have dominate population while $\pm 2\hbar k_L$ and $0\hbar k_L$ for the S-band, as indicated by the experimental results shown in Fig.~\ref{shortcut_method}(b). Due to the symmetry of parity, two series of pulses $V_{0} \cos ^{2}\left(k_L x\right)$ and $V_{0}\sin ^{2}\left(k_L x\right)$ are needed if we want to load directly a BEC into the F-band. For convenience, we load atoms into the G-band with $V_{0} \cos ^{2}\left(k_L x\right)$ pulses sequence in the experiment.

Our loading process is finished in tens of microseconds by applying a series of pulsed optical lattices, and this rapid generation of the quantum state in higher bands is different from other preparation methods~\cite{Wirth1,M}. Our method allows us to explore rich physics in high Bloch bands predicted theoretically~\cite{IBloch2,NP,Speed,Shaken,Wu,Liu1,Larson}.

\section{Experimental observation}
\begin{figure*}
\begin{center}
  \includegraphics[width=14.5cm]{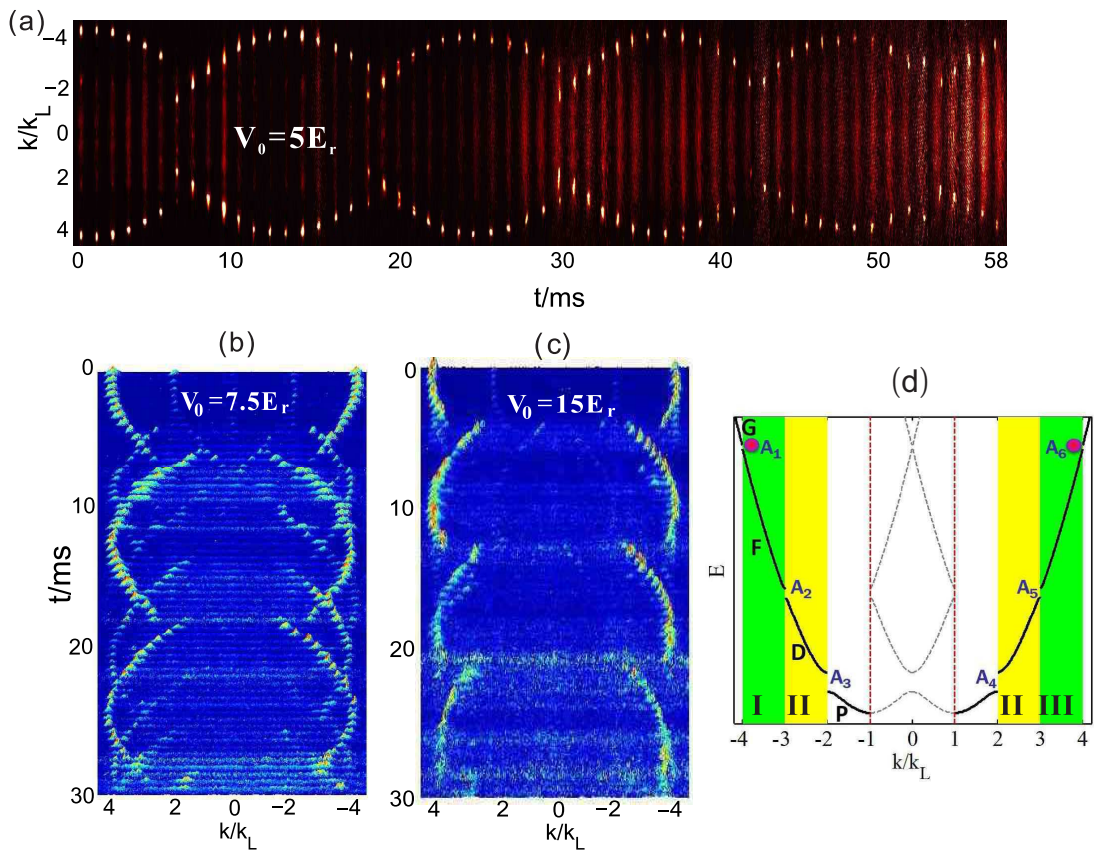}
\end{center}
  \caption{(Color online) Quantum oscillations of the BEC in high bands of optical lattices. Shown in (a), (b) and (c) are experimental results in momentum space with lattice depth $V_0=5E_r$, $V_0=7.5E_r$ and $15E_r$, respectively. The oscillations in (a) are across both the F and D bands and oscillations in (c) are only within the F-band; the dynamics in (b) are superposition of the oscillations in (a) and (c). (d) is the schematic of extended Bloch bands (P, D, F and G) of a one-dimensional optical lattice. The green areas (I and III) are for the F-band and the yellow (II) for the D-band.}\label{Experiment}
\end{figure*}

After the BEC in the G-band is prepared, we hold the lattice and harmonic trap for a period of time $t$, and then switch off all potentials to take pictures after $28$ms time of flight (TOF). Three series of experimental absorption images for the optical lattice depth $V_0=5E_r$, $7.5E_r$, and $15E_r$ are shown, respectively, in Fig.~\ref{Experiment}(a,b,c). $E_r=\hbar^2 k_L^2/2m$ is the recoil energy with $m$ being the atomic mass. The time separation between neighboring images in the $5E_r$ series is $1$ms and $0.5$ms in the other two series. These series of images demonstrate clearly three different quantum oscillations; we will explain and analyze them later. For convenience, an extended band structure is drawn in Fig.~\ref{Experiment}(d), where the energy gaps between different bands are marked with $A_s$ ($s=1,2,3,4,5,6$).

The BEC is initially loaded in the G-band, where the atoms mostly populate equally around
two momenta $\pm 4\hbar k_L$. As atoms with these two momenta are at the center of the
trapping potential at the beginning, the only possible motion for them is to move either to the left or the right. Consequently, they lose their momenta while gaining harmonic potential energy. This corresponds to that the BEC makes a quantum tunneling from the G-band to the F-band over the tiny band gap at $A_1$ and $A_6$ in Fig.~\ref{Experiment}(d). It is impossible for the BEC to move up along the G-band due to the conservation of energy. When the lattice depth is very high so that the energy gap at $A_1$ and $A_6$ is large, it is possible for the BEC to stay in the G-band for a long time. We have observed atoms maintaining in the G-band for $5$ms when $V_0=20E_r$. For the lattice strength of interest in our experiment, the BEC always tunnels  from the G-band to the F-band as soon as the initial loading ends.

Once the BEC is in the F-band, it continues to lose momentum while gaining harmonic potential energy. This corresponds to that the BEC traverses dynamically along the F-band from $A_1$, $A_6$ to $A_2$, $A_5$ in Fig.~\ref{Experiment}(d). Once arriving at $A_2$ and $A_5$, the atoms face different ensuing dynamics depending on the lattice strength. If the lattice strength is small and the Bragg reflection at $A_2$ and $A_5$ is weak, the BEC will continue its dynamics into the D-band by crossing the band gap. After evolving dynamically along the entire D-band, the BEC comes to the band gap between D and P bands at $A_3$ and $A_4$. This band gap is always large for the lattice strength in our experiments. As a result, the atoms at $A_3$ ($-2\hbar k_L$) will be Bragg reflected completely to $A_4$ ($2\hbar k_L$) while the atoms at $A_4$ will be Bragg reflected completely to $A_3$. No tunneling to P-band occurs. Afterwards the BEC will reverse its dynamics by moving up in momentum from $A_4$, $A_3$ to $A_5$, $A_2$. It eventually arrives at $A_6$, $A_1$, finishing half of an oscillating cycle. These oscillations crossing the two Bloch bands (F and D bands) are driven under a variable force from the harmonic trap. They are illustrated in Fig.~\ref{Experiment}(a) for lattice depth $V_0=5E_r$ and their period is $24$ms.

Note that the BEC moving up along the bands around $A_4$, $A_3$ in Fig.~\ref{Experiment}(d) by gaining momenta is due to the fact that most of the atoms are away from the center of the trapping potential and feel an accelerating force. This is different from the initial stage when the BEC is loaded into the G-band, where most of the atoms are at the center of the trap and feel very small force for the finite size of the BEC.

When the optical lattice is strong and the gap at $A_2$, $A_5$ is large, the Bragg reflection can dominate the dynamics, forbidding the atoms tunnel from the F-band to the D-band. Instead, the atoms at $A_2$ ($-3\hbar k_L$) will transfer completely to $A_5$ ($3\hbar k_L$) via Bragg reflection while the atoms at $A_5$ will also transfer completely to $A_2$. In this way, the quantum dynamics is confined within the F-band. These oscillations only within the F-band are observed in our experiment for $V_0=15E_r$
and are shown Fig.~\ref{Experiment}(c) with a period of $17$ms.

\begin{figure}
\begin{center}
  \includegraphics[width=0.45\textwidth]{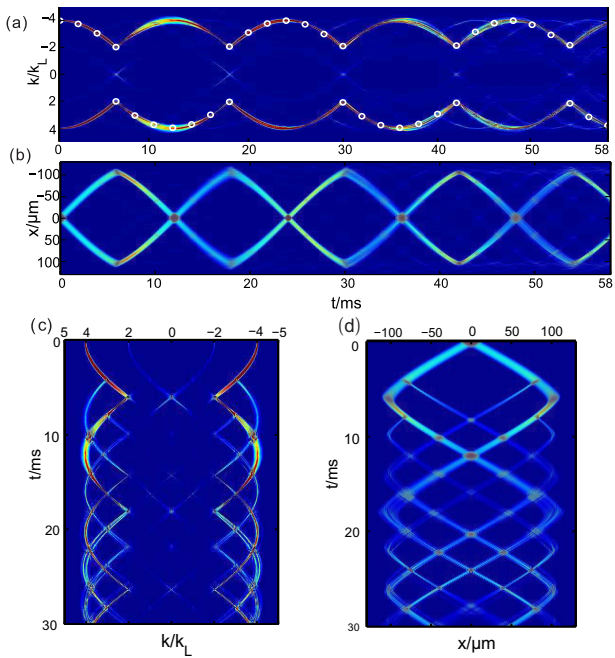}
\end{center}
  \caption{(Color online) The theoretical results of quantum dynamical oscillations for the BEC in high bands in both real space and momentum space, respectively. Shown in (a) and (b) are results for $V_0=5E_r$ while (c) and (d) are $V_0=7.5E_r$. The white circles in (a) are the result of the semi-classical model.}\label{Theory}
\end{figure}

When the lattice strength is intermediate, the Bragg reflection at $A_2$, $A_5$ will be partial: one part of the atoms will be reflected and undergo oscillations within the F-band; the other part of the atoms will tunnel to the D-band and oscillate across both the F and D bands. As a result, we should be able to observe a superposition of the two kinds of oscillations: across both the F and D bands and only within the F-band, when the lattice is at an intermediate strength. This is indeed what we observed in experiment as shown in Fig.~\ref{Experiment}(b) and simulated in theory as shown in Fig.~\ref{Theory}(c) and (d) for $V_0=7.5E_r$, where the two kinds of oscillations are clearly seen, and the ratio of them can be tuned by the lattice strength.

%The non-adiabatic loading into the high bands has some heating effect. However, it is not too severe. Our measurement shows that around 60\% percent of the cloud is still a BEC. After 28 ms of flight, almost all the thermal cloud is dispersed due to their high velocities. The TOF images in Fig.1 are almost entirely from the BEC part.

\section{The theoretical explanation and simulation}
We have simulated the experiment with the one-dimensional Gross-Pitaevskii equation (GPE).  In the simulation, the initial state is $\psi \left( {x,t = 0} \right) = {\psi _g}\left( x \right)\phi_{q=0}(x)$, where $\psi_g(x)$ is the ground state of the BEC in a harmonic trap and $\phi_{q=0}(x)$ is the Bloch wave function in the G-band at $q=0$. We have plotted our numerical results in both the real space and the momentum space in Fig.~\ref{Theory} for $V_0=5 E_r$ and $V_0=7.5 E_r$. The results in the momentum space agree well with the experimental results in Fig.~\ref{Experiment}. We can infer from Fig.~\ref{Theory}(b) that the BEC can move away from the center of the trap by up to $\pm 100 \mu$m for the oscillations in both the F and D bands. For oscillations in F-band, our numerical results indicate that  this displacement can be up to $\pm 75 \mu$m, which is much larger than what was observed in super-Bloch oscillations in Ref.~\cite{Haller}. There is a small discrepancy between the theoretical results and the experimental results. For example, there is a slight off-set in the experimental oscillations around 6-7ms for $V_0=5 E_r$; this slight off-set is not seen in the corresponding theoretical results. The explanation for this small discrepancy becomes clear when we discuss Brag reflection in the next section.

It is difficult to observe oscillations in the real space in experiment. To do that, we need to keep the TOF very short. Within such a short TOF, the atomic cloud is still very dense and the TOF image is not proportional to the cloud density. At the same time, the TOF image is obscured by the thermal cloud that has no time to disperse.

As the BEC density profile varies smoothly over hundreds of lattice sites, its dynamics can be well described by the semi-classical dynamics of Bloch particles~\cite{Mermin},
\begin{eqnarray}
\label{re}\hbar\frac{d{\bm r}}{dt}&=&{\nabla_{\bm q}} E_n({\bm q})\,\\
\label{qe}\hbar\frac{d{\bm q}}{dt}&=&{\bm f}({\bm r})\,
\end{eqnarray}
where $E_n({\bm q})$ is the $n$th energy band and ${\bm f}({\bm r})$ is the force acting on the Bloch particle. For Bloch electrons in traditional condensed matter physics, we usually have ${\bm f}({\bm r})=-e {\bm E}-e\frac{d{\bm r}}{dt}\times {\bm B}$. In our case, ${\bm f}({\bm r})=-m\omega_x^2 x {\hat e_x}$ with the harmonic trap frequency $\omega_x$.

For simplicity, we describe the energy bands with a cosine function as ${E_n}\left( q \right) = {A_n} + \frac{{{B_n}}}{2}\cos \left( {q \pi/k_L} \right)$, with $n=\rm F,\rm D$ representing the F and D bands, respectively, and $\left| {B_n} \right|$
the width of the energy bands. For this simplified case, we can solve Eqs.(\ref{re},\ref{qe}) analytically and find that the oscillation periods are:
\begin{eqnarray}\label{eqn:T_F}
T_{\rm F} = \frac{4\hbar k_L}{\sqrt{m} \omega _x \pi} \frac{C_1}{\sqrt{B_{\rm F}}}
\end{eqnarray}
and
\begin{eqnarray}\label{eqn:T_FD}
T_{\rm FD} =\frac{4\hbar k_L}{\sqrt{m} \omega _x \pi} \left( \frac{C_1}{\sqrt{B_{\rm F}}} +\frac{C_2}{\sqrt{- B_{\rm D}}} \right),
\end{eqnarray}
where $C_1= \int_{{q_ + }}^{{k_L}} {\frac{{\rm d}\left( {qa} \right)}{\sqrt {\cos \left( {{q_ + }a} \right) - \cos \left( {qa} \right)} }}$ and $C_2= u \textbf{K}\left( u \right)$. $q_ +$ is the initial quasi-momentum considering the finite size of the BEC, $u = \left( {\frac{B_F}{ - {B_D}}{{\cos }^2}\frac{{q_ + }a}{2} + 1} \right)^{-1/2}$, and $\textbf{K}\left( u \right)$ is the complete elliptic integral of the first kind as ${\mathbf{K}}\left( u \right) = \frac{\pi }{2}\left\{ {1 + \sum\limits_{n = 1}^{ + \infty } {{{\left[ {\frac{{\left( {2n - 1} \right)!!}}{{{2^n}n!}}} \right]}^2}{u^{2n}}} } \right\}$. It is obvious that the periods for the two kinds of oscillations are inversely proportional to $\omega_x$.

\begin{figure}
\begin{center}
  \includegraphics[width=0.35\textwidth]{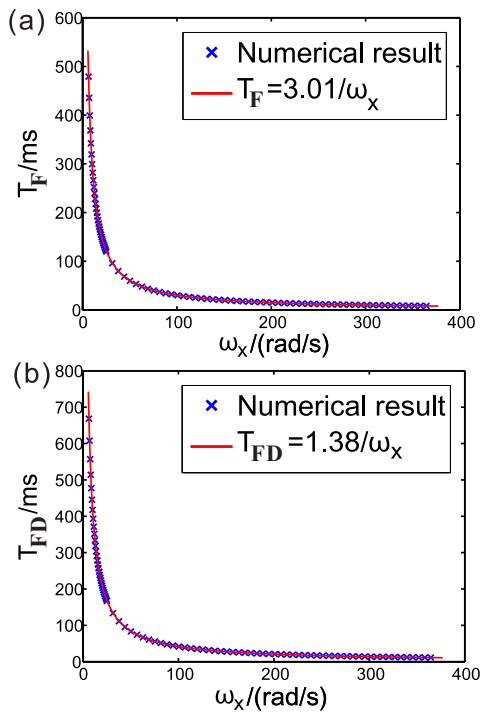}
\end{center}
  \caption{(Color online) (a) The periods of the two kinds of oscillations for: $T_{\rm F}$ for $V_0=15E_r$; (b) $T_{\rm FD}$ for $5E_r$. The red lines are fitting functions: (a) $T_{\rm F}=3.01/ \omega_x$; (b) $T_{\rm FD}=1.38/ \omega_x$.}\label{periods}
\end{figure}

This inverse relation does not change even when we use the realistic Bloch bands instead of the idealized cosine form. This is confirmed by our numerical results with the semi-classical equations (\ref{re},\ref{qe}) as shown in Fig.~\ref{periods}. Since $\omega_x=2 \pi \times 28$Hz in our experiment, we have $T_{\rm F}=17.1$ms for $V_0=15E_r$ and $T_{\rm FD}=23.9$ms for $V_0=5E_r$, which agree very well with the experimental results in Fig.~\ref{Experiment}. In Fig.~\ref{Theory}(a), the semi-classical oscillations are plotted as white circles, matching both the experimental result and the numerical GPE result.\\

\section{High order Bragg reflection}
\begin{figure}
\begin{center}
  \includegraphics[width=0.4\textwidth]{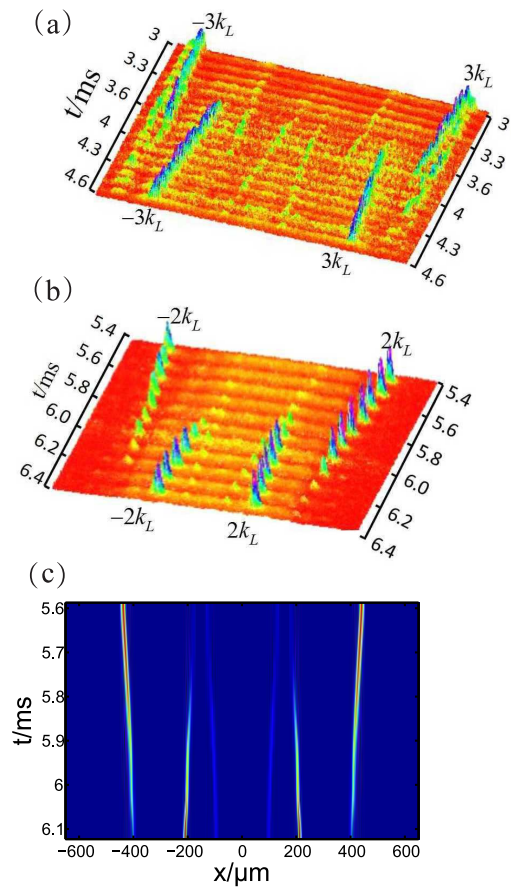}
\end{center}
  \caption{(Color online) High order Bragg reflections process observed (a) at the edge of the F-band for $V_{0}=15E_r$; (b) at the center of the D-band for $V_{0}=5E_r$ in the FBZ. (c) is the simulation results from GPE for $V_0=5E_r$.}\label{Bragg_reflections}
\end{figure}

Bragg reflection (or scattering) at the Bloch band edge or center is a fundamental quantum process in periodic systems. As we have already seen, it plays a crucial role in the oscillations observed in our experiments. We now take a closer look at it by recording the absorption images every $0.1$ms. Two series of images are shown in Fig.~\ref{Bragg_reflections}, where we see clearly the reflection process between $\pm 3\hbar k_L$ in (a) under $15E_r$ and between $\pm 2\hbar k_L$ in (b) under $5E_r$. Bragg reflections have been demonstrated in lower bands with ultracold~\cite{Salomon,Kozuma,Zhang} atomic gases, and in high bands with ultracold atoms by using a time-dependent optical lattice in Ref.~\cite{Park}, where only reflection results were given, but the reflection processes weren't shown. To the best of our knowledge, we are the first to directly observe high order Bragg reflections process in high excited bands of optical lattices.

We use Fig.~\ref{Bragg_reflections}(b) to show in detail what is observed in our experiment. The Bragg reflection occurs roughly between 5.8ms and 6.1ms. During this period, atoms around $2\hbar k_L$ get scattered to $-2\hbar k_L$ while atoms around $-2\hbar k_L$ get scattered to $2\hbar k_L$. As a result, there are two fractions of atoms at $2\hbar k_L$: one fraction waits to be reflected to $-2\hbar k_L$ and the other consists of atoms just scattered from $-2\hbar k_L$. As atoms at $\pm 2\hbar k_L$ are located in different places, these two fractions at $2\hbar k_L$ are separated in coordinate space. With a 28ms TOF, these two fractions appear as two different peaks. There are also two fractions for $-2\hbar k_L$. As a result, we observe four peaks during the period between 5.8ms and 6.1ms, instead of two peaks for other times.

As shown in Fig.~\ref{Bragg_reflections}(a), the reflection between $\pm 3\hbar k_L$ occurs roughly from 3.4ms to 4.0ms, which is about twice longer than the reflection between $\pm 2\hbar k_L$. This shows that higher-order Bragg reflection is more difficult to occur. Fig.~\ref{Bragg_reflections}(c) is the simulation result with the GPE for $5E_r$ which is consistent with our experimental result.

\begin{figure}
\begin{center}
  \includegraphics[width=0.4\textwidth]{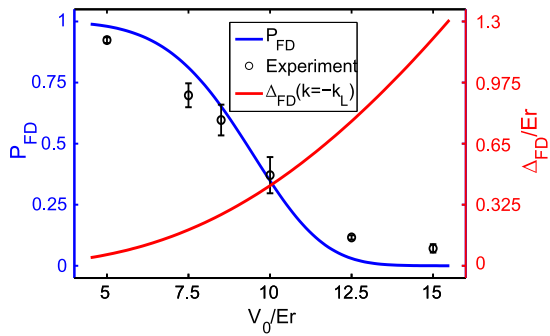}
\end{center}
  \caption{(Color online) The calculated transition probability from the F-band to D-band versus $V_0$ is shown as blue solid line. The black circles are the experimental results. The band gaps between the F and D bands at the FBZ edge versus the lattice depth are also shown as red solid line.}\label{transition_probability}
\end{figure}

When the atoms are not Bragg reflected, they tunnel from one band to a neighboring band. This is the well known LZ tunneling. In other words, the Bragg reflection can be described as a complementary process to the LZ tunneling. Between the F and D bands, the tunneling is determined by the $P_{\text{FD}}={e^{-2\pi \gamma_{\text{FD}} }}$~\cite{Zener} with
\begin{eqnarray}\label{eqn:Probability}
\gamma_{\text{FD}} = \frac{\Delta^2_{\text{FD}}}{4\hbar} {\left| \frac{6 \hbar k_L f(x)}{m} \right|^{ - 1}}.
\end{eqnarray}
The calculated probability $P_{\text{FD}}$ versus $V_0$ as blue solid line is shown in Fig.~\ref{transition_probability}. It is clear that the tunneling probability decreases with the increasing lattice depth. It is close to $1$ at $V_0=5E_r$ while close to zero at $15E_r$. Black circles are experimental results for the ratio of atoms tunneled to the D-band. There is a very good agreement between the experiment and the theory. The band gaps $\Delta_{\text{FD}}$ between the F and D bands at quasi-momenta being $\pm \hbar k_L$ versus $V_0$ are shown as red solid line in Fig.~\ref{transition_probability} to assist the understanding of this quantum transition.

\section{Conclusion}
In summary, we have loaded a BEC in a one-dimensional optical lattice non-adiabatically into the G-band. The BEC is then observed to tunnel to the F-band and undergoes quantum oscillations within the F and D bands. The variable force exerted on atoms enables a direct visualization of oscillations in high bands in the momentum space. By controlling the Bragg reflection at the edges of the F and D bands with optical lattice, we have observed three different types of quantum oscillations. At weak lattice strength, oscillations between both the F and D bands are observed; at strong lattice strength, oscillations only within the F-band are observed. At intermediate strength, a superposition of the above two dynamical oscillations is observed. Furthermore, we has directly demonstrated the high order Bragg reflections process in high excited bands of optical lattices.

\section*{ACKNOWLEDGEMENTS}

This work is supported by the state Key Development Program for Basic Research of China NSFC (Grants No.61475007, No.11334001 and No.91336103).

\bigskip
\emph{}

\end{document}